\documentclass[twocolumn]{article}

\usepackage{amsmath,amssymb, fullpage}
\usepackage[pdftex]{color}
\usepackage{graphicx}

\DeclareGraphicsExtensions{.pdf,.jpg}

\newcommand{\eone}[1]{\langle #1 \rangle}

\begin{document}

\title{Probing Phases and Quantum Criticality using Deviations from the Local Fluctuation-Dissipation Theorem}

\author{ E. Duchon$^1$, Y. Kato$^2$, N. Trivedi$^1$\\
$^1$Department of Physics, Ohio State University, Columbus, Ohio
43210, USA\\
$^2$Theoretical Division, Los Alamos National Laboratory, Los Alamos, New Mexico 87545 USA
}

\maketitle

{\bf Introduction
Cold atomic gases in optical lattices are emerging as excellent laboratories for testing models of strongly interacting particles in condensed matter physics. 
Currently, one of the major open questions is how to obtain the finite temperature phase diagram of a given quantum Hamiltonian directly from experiments. Previous work in this direction required quantum Monte Carlo simulations to directly model the experimental situation in order to extract quantitative information, clearly defeating the purpose of an optical lattice emulator. 
Here we propose a new method that utilizes deviations from a local fluctuation dissipation theorem to construct a finite temperature phase diagram, for the first time, from local observables accessible by {\it in situ} experimental observations. Our approach extends the utility of the fluctuation-dissipation theorem from thermometry to the identification of quantum phases, associated energy scales and the quantum critical region. We test our ideas using state-of-the-art large-scale quantum Monte Carlo simulations of the two-dimensional Bose Hubbard model.
}

\begin{figure*}[htb!]
\begin{center}
\vspace{-0.3cm}
\includegraphics[width=0.85\textwidth]{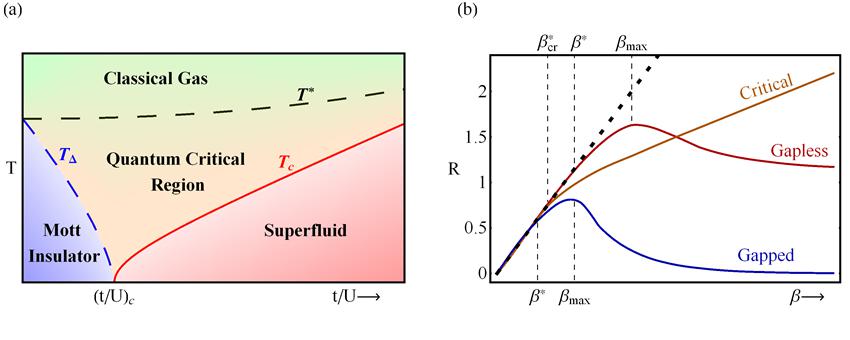}
\end{center}
\vspace{-0.75cm}
\caption{Identifying Phases and the Quantum Critical Region using Deviations from the Fluctuation Dissipation Theorem:
(a) Finite temperature phase diagram of a quantum Hamiltonian, e.g. the Bose Hubbard model, shows the suppression of the superfluid transition temperature $T_c$ upon repulsive interaction $U$ increasing relative to tunneling $t$ and ultimately vanishing at a critical coupling $(t/U)_c$. 
For $t/U<(t/U)_c$, the system is Mott insulating with a crossover scale $T_{\Delta}$ that also vanishes at $(t/U)_c$. $T^\ast$ is the temperature scale at which local quantum effects begin to emerge. The shaded region above $(t/U)_c$ depicts the quantum critical region dominated by large fluctuations. 
(b) We show how such a phase diagram for a given quantum Hamiltonian can be obtained from the local fluctuation-dissipation ratio $R$ defined in equation~(\ref{definitionR}). In the gapped Mott phase we extract $\beta_{\rm max}$, which is related to the inverse $T_\Delta$ as discussed in the text; similarly, the maximum in the gapless SF phase $\beta_{\rm max}$ corresponds to the inverse $T_c$. For both phases the deviation of $R$ from linearity marks 
$\beta^\ast=1/T^\ast$ where quantum effects begin. In the critical region, $R$ remains linear in $\beta$ but with a proportionality constant less than unity.}
\label{qpt}
\end{figure*}

As thermal fluctuations decrease, interactions between particles dominate and drive systems into emergent quantum phases of matter such as superfluids, Mott insulators, magnetically ordered phases, and spin liquids\cite{greiner02,anderson04,highTc06,yan11}. Not only are these phases with their associated excitations interesting, but the transition between quantum phases opens up a quantum critical region dominated by large fluctuations. The large fluctuations arise from new degrees of freedom that must form as the system transits from one phase to the other, tuned by a coupling parameter as illustrated in Fig.~\ref{qpt}.

Quantum gases confined in an optical lattice offer a unique platform to study such quantum phase transitions since the coupling strength is easily tuned by the laser intensity or the magnetic field. Detection and characterization of these systems has, until recently, been limited to time-of-flight observations. Experiments on bosons in optical lattices, emulating the Bose Hubbard model (BHM), have shown signatures of the coherent superfluid (sharp peaks) and of the Mott state (broad, featureless distributions) \cite{greiner02,spielman08}. Further theoretical investigation has indicated that sharp peaks in the momentum distribution can occur even in the non-superfluid phase and there is a wealth of information about quantum critical fluctuations in the nature of these peaks \cite{kato08,diener07}.

The primary bottleneck in obtaining a more quantitative comparison with theory has proven to be a definitive measurement of the temperature and a diagnostic of the all-important quantum critical regime. Some preliminary identifications have been made, but they require input from individually tailored large-scale quantum Monte Carlo (QMC) simulations \cite{trotzky10,fang11}. 

In this Letter, we provide the theoretical framework for the application of a local fluctuation dissipation (LFD) theorem to experiments to gain fundamental insights into the nature of the phases, their low-lying excitations and quantum criticality. 
Motivated by the success of recent experiments\cite{bakr09, sherson10, hung11,Bakr10,Gemelke09} to access the local density {\it in situ}, we define a LFD ratio $R$ as the ratio of the local compressibility $\kappa_i$ to local number fluctuations $\delta n_i^2$ at site $i$, defined by
\begin{eqnarray}
R&=&\kappa_i/\delta n_i^2  \\ \nonumber
\kappa_i&=&\partial \langle n_i\rangle /\partial \mu \\ \nonumber
\delta n_i^2&=&\langle n_i^2\rangle -\langle n_i\rangle ^2
\label{definitionR}
\end{eqnarray}

Remarkably, from the temperature dependence of this single quantity $R$, we propose that it is possible 
to estimate the temperature of the onset of quantum effects  $T^*$, the temperature of quantum phase ordering $T_{max}$ and the quantum critical region, for a general quantum Hamiltonian. 
We test our proposal on the 2D BHM using large-scale QMC simulations and summarize our central results for phase identification in Fig.~\ref{qpt}(b) and in the finite-temperature phase diagram in Fig.~\ref{phasediagram}, obtained solely from this LFD ratio. 
This explicit demonstration 
opens up the possibility of finding such phase diagrams for general quantum Hamiltonians {\em directly} from experimental data. 
We thus provide the crucial missing link in the grand challenge to emulate strongly correlated materials such as the high temperature superconductors using ultracold atoms in optical lattices.

\noindent {\bf Fluctuation-Dissipation Theorem (FDT):} The full quantum FDT relates the imaginary response function $\chi''$ to the dynamic structure factor $S$ at inverse temperature $\beta$ and is given by
\begin{equation}
\chi^{\prime\prime}({\bf q},\omega)=\frac{1-e^{-\beta\omega}}{2}S({\bf q},\omega)\  \cdot
\end{equation}
Here we specialize to perturbations of the density $\rho({\bf q})=\sum_{\bf k} a_{{\bf k}+{\bf q}}^\dagger a_{\bf k}$ and the density-density correlation function 
$S({\bf q},t)=\langle \rho({\bf q},t)\rho({\bf 0},0)\rangle$. 
A conserved quantity, such as the total number of particles $N=\rho({\bf q}=0)$, commutes with the Hamiltonian, so $\rho({\bf q}=0,t)$ is independent of $t$. As discussed in the supplement, this leads to the exact expression 
$\chi({\bf q}\rightarrow0,\omega=0)=\beta\; S({\bf q}\rightarrow0,t=0)$. The zero frequency response function equals $\partial n/\partial \mu\equiv n^2\kappa_T$ where $\kappa_T$ is the thermodynamic compressibility, 
and the equal-time correlation function equals $\beta\; \delta N^2 /V$ where $\delta N^2=\eone{N^2}-\eone{N}^2$ is the total number fluctuations.  Equating the two gives 

\begin{equation}
n^2\kappa_{T}= \frac{\partial \eone{n}}{\partial\mu}= \beta \frac{\delta N^2}{V}\  \cdot
\label{fullFDT}
\end{equation}
While equation~(\ref{fullFDT})  resembles the classical FDT, it is crucial to note that it is valid even in the quantum regime since $\rho(\bf{q}=0)$ is a conserved quantity.

There are two ways to probe the system locally. We could locally perturb the system with a small change in the chemical potential $\delta\mu_i$ at site $i$ and measure the resulting local density variation $\delta n_i$.
This procedure leads to the single site form of the classical FDT,
$
\kappa_{i\;L}\equiv\frac{\partial \eone{n_i}}{\partial\mu_i}=\beta\;\delta n_i^2,
$
that is only valid at high temperatures since the density on a given site is not conserved. Both forms of the FDT relations in equation~(\ref{fullFDT}) and in the single site form $\kappa_{i\;L}$ discussed above enable thermometry \cite{zhou09b,ma10,fang11}, but serve a limited purpose in revealing other properties of the system. 

We propose that a more useful quantity, sensitive to the nature of the phase and to various energy scales, is the local fluctuation-dissipation (LFD) ratio $R$ defined in equation~(\ref{definitionR}). 
The essential difference is that $R$ involves $\kappa_i=\partial \langle n_i \rangle /\partial \mu$ 
which is the measured change of the {\em local} density in response to a {\em global} chemical potential $\mu$ variation, and as such is 
sensitive to long range order and phase transitions, as opposed to $\kappa_{i\;L}$ which is the local response to a local perturbation. 
For a homogeneous system, $\kappa_i$ is the same as the thermodynamic compressibility (up to factors of $n^2$). 
Although we investigate $R$ in a uniform system below, it is important to keep in mind the applicability to inhomogeneous systems as created in trapped atomic gases. Assuming local density approximation, $\kappa_i$ and $R$ can be extracted directly from the density profile $n(\mu(r))$ \cite{zhou09a}. 

\begin{figure}[t!]
\begin{center}
\includegraphics[width=0.4\textwidth]{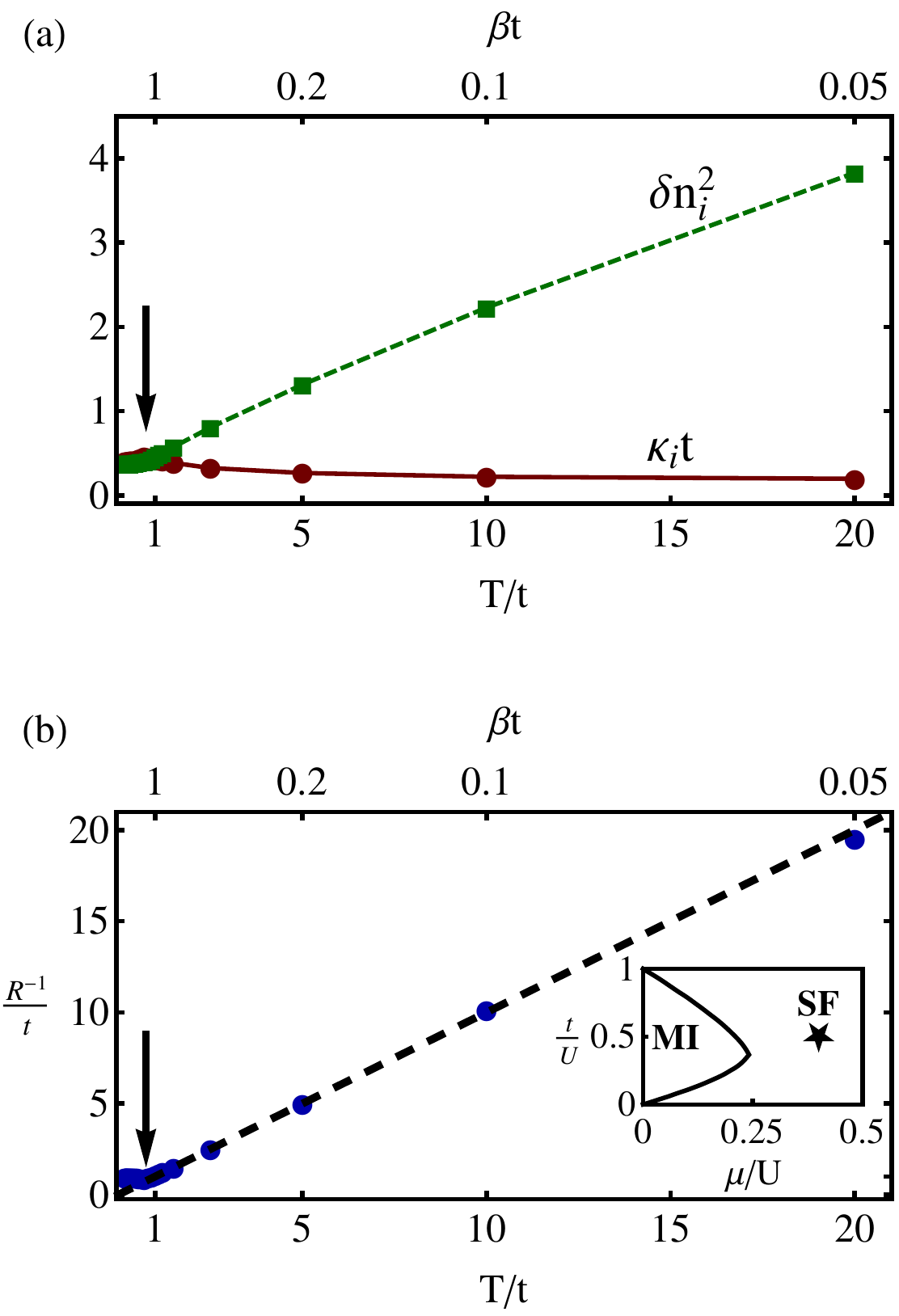}
\vspace{-0.7cm}
\end{center}
\caption{High Temperature Behavior of Local Observables. (a) The local compressibility $\kappa_i=\partial \eone{n_i}/\partial\mu$ and the local number fluctuations $\delta n_i^2$ well inside the SF phase of the 2D BHM. As expected for an ideal gas, $\kappa_i$ approaches a constant and $\delta n_i^2$ grows with $T$ at high temperatures. (b) $R^{-1}=\delta n_i^2/\kappa_i \approx T$ confirms the high temperature behavior (dashed line) of the LFD ratio, equation~(\ref{definitionR}). Deviation of $R^{-1}$ from linear behavior allows us to determine the onset of quantum effects.
\label{highT}
}
\end{figure}

In the following, we demonstrate the potential for $R$ to map out finite temperature phase diagrams by evaluating $R$ in the BHM. Bosons trapped in an optical lattice and confined in a potential are described by 
\begin{align}
H = &-\frac{t}{z} \displaystyle\sum_{\langle i,j\rangle}(a_i^{\dagger}a_j+a_i a_j^{\dagger}) \nonumber \\ 
&+\frac{U}{2}\displaystyle\sum_i n_i(n_i-1) - \displaystyle\sum_i \mu_i n_i.
\end{align}
Here $a_i\;(a_i^{\dagger})$ are boson annihilation (creation) operators, $\langle\ldots\rangle$ indicates nearest neighbor sites, $\mu_i=\mu_0-\alpha r_i^2$ is the chemical potential on site $i$ for a parabolic confining potential, and $t$ and $U$ set the hopping and interaction energy scales, respectively. We simulate the BHM at finite temperatures with worldline QMC using the directed loop algorithm on up to $32^2$ site lattices \cite{kato09}. We establish the essential ideas in a uniform system $(\alpha=0)$ for clarity, but the results are easily extended to the nonuniform system by using the local density approximation.

\noindent {\bf Classical Regime:} At high temperatures, $T\gg U \gg t$, the system is in a non-interacting classical regime. 
Here the inverse LFD ratio $R^{-1}\approx T$ is determined primarily by the density fluctuations $\delta n_i^2$ that increase linearly with temperature 
($\kappa_i$ remains finite and independent of $T$).  As $T$ decreases to the regime $U\gg T\gg t$, the system remains classical. Interaction effects cause both $\kappa_i$ and $\delta n_i^2$ to deviate from the high-T ideal gas limit but 
the inverse LFD ratio remains approximately linear in $T$ (see Fig.~\ref{highT}).

$R$ can also be used to test for equilibration of the system in different regions. For bosons in optical lattices, the Mott-like center is sometimes observed to be at a significantly different temperature from the superfluid or normal wings~\cite{hung10}. In such situations, successive local measurements of $R$ can be useful to garner information about rate-limiting processes for achieving equilibrium.

\begin{figure*}[t!]
\begin{center}
\includegraphics[width=0.95\textwidth]{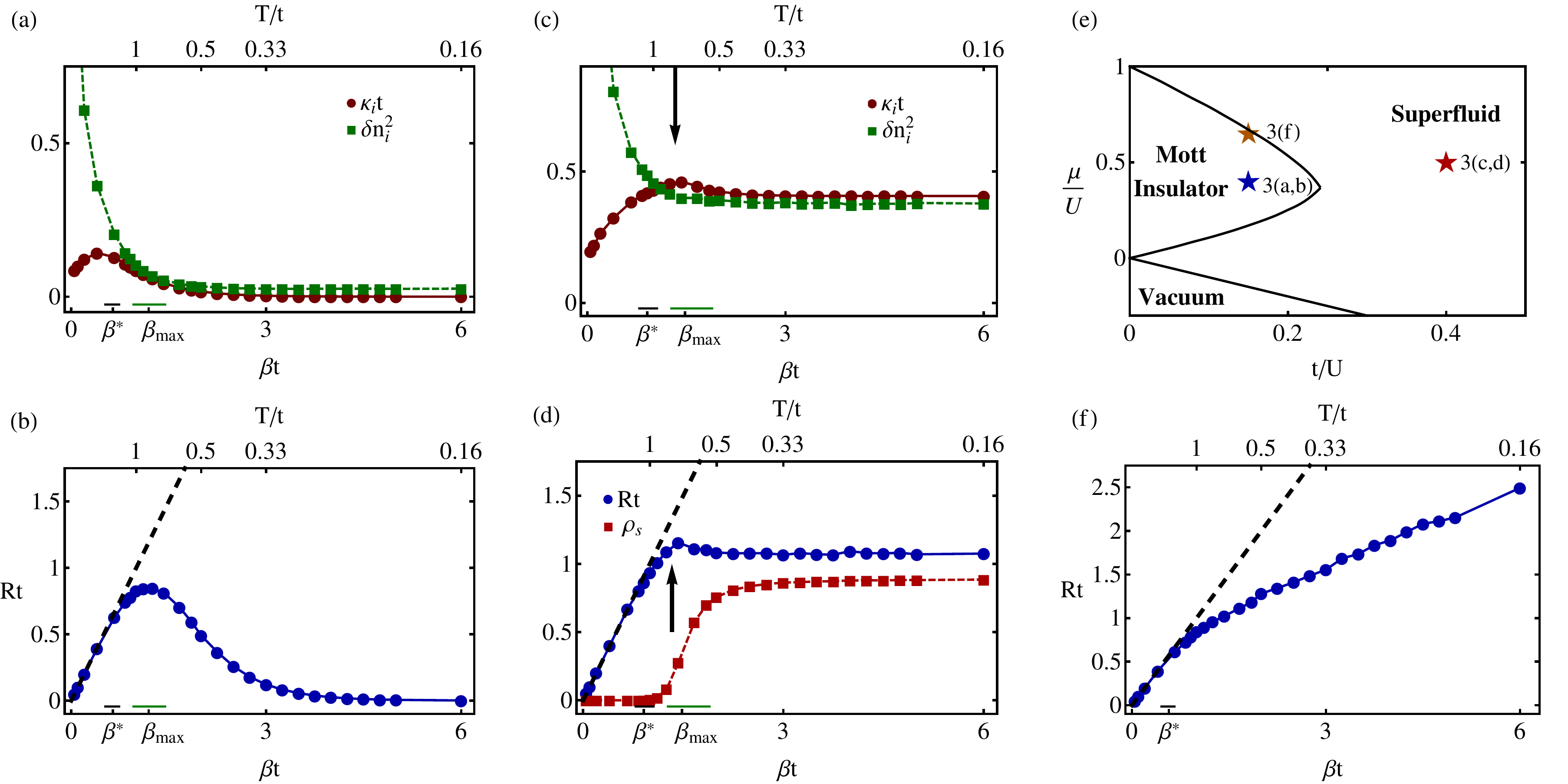}
\end{center}
\vspace{-0.6cm}
\caption{\label{phasesignatures} Phase Signatures and Energy Scales Encoded in $R$.
Panel (e) The $T=0$ BHM phase diagram with the parameters of each dataset are indicated.
Panels (a) and (b): In the MI, the compressibility $\kappa_i$ (circles) is suppressed at low temperatures to zero while local quantum fluctuations maintain $\delta n_i^2$ (squares) at a finite value. 
The corresponding LFD $R=\kappa_i/\delta n_i^2$ in the MI shows a characteristic maximum and decays exponentially at low $T$, 
from which we determine the particle-hole gap energy $\Delta_{ph}$ discussed in the Supplement.
Panels (c) and (d): In the SF, $\kappa_i$, $\delta n_i^2$ and $R$ all approach a finite constant as $T\rightarrow0$. 
The peak in $R$ occurs at $\beta_{max}$ and agrees well with the inverse superfluid $T_c$ (indicated by the arrow) 
obtained from the superfluid density $\rho_s$ (squares in (d)). 
In all panels (a,b,c,d), the inverse temperature $\beta^*$ indicates when $R$ deviates significantly from the classical limit. 
The dashed black line indicates the high temperature $R\approx\beta$ limit in (b,d) and horizontal bars indicate uncertainty in $\beta^*$ and $\beta_{max}$ in each panel.
Panel (f) shows the behavior close to the quantum critical point, where the system does not order at the temperatures probed. Both $\delta n_i^2$ and 
$\kappa_i$ (not shown) do not converge to the behavior of either phase, and $R\propto\beta$ with a slope less than one indicates the absence of an energy scale, a characteristic of the quantum critical region. 
}
\end{figure*}

\begin{figure*}[t!]
\begin{center}
\includegraphics[width=1.0\textwidth]{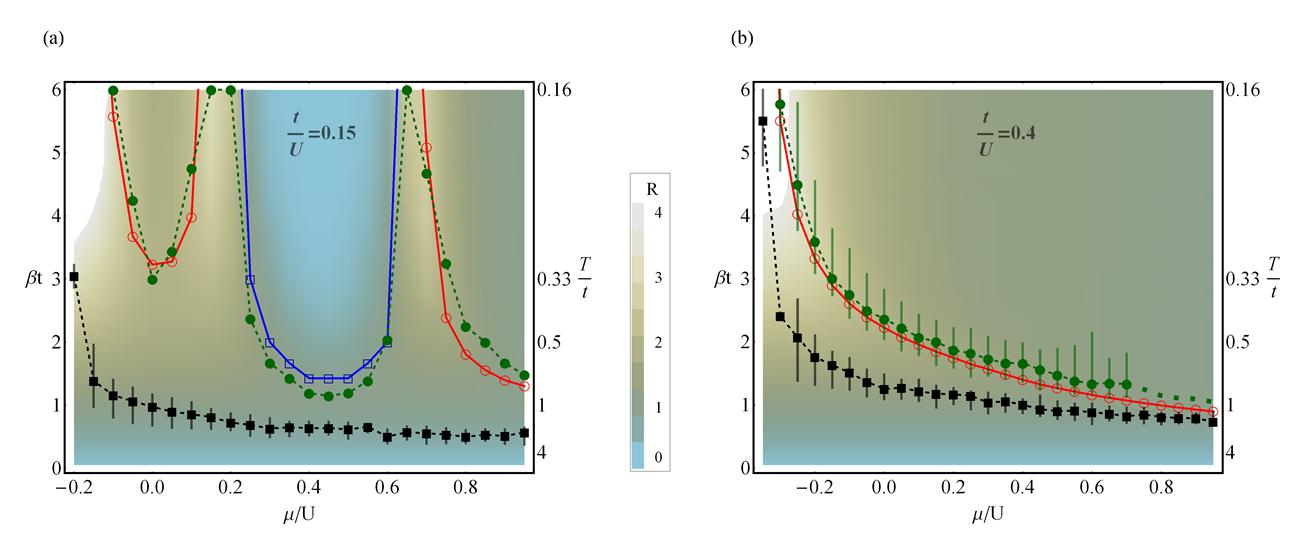}
\end{center}
\vspace{-0.75cm}
\caption{\label{phasediagram} Determination of the Phase Diagram of a Quantum Hamiltonian using Local Observables: 
This figure shows the contour plot of LFD ratio $R$ in the inverse temperature-chemical potential ($\beta t-\mu/U$) plane at (a) $t/U=0.15$ and (b) $t/U=0.4$. 
We determine the phase diagram from $R(\beta)$ by marking two inverse temperature scales, $\beta^*=1/T^*$ (solid squares) that indicates the onset of quantum effects, and $\beta_{max}=1/T_{max}$ (solid circles) that delineate phase boundaries or crossovers. 
$\beta_{max}$ agrees with the characteristic temperature scales in the superfluid and the MI phases, or $1/T_c$ (open circles) and $1/T_{\Delta}$ (open squares), respectively.
Here $T_c$ and $T_\Delta$ are calculated independently from R by using the vanishing of $\rho_s$ and $\kappa_i$ and confirm the interpretation of the $T_{max}$ scale.
As seen from Fig.~\ref{phasesignatures}(a), at $T=0$ and $t/U=0.15$, the system progresses from vacuum to SF, hits two quantum critical points bounding the MI, and returns to SF as $\mu/U$ increases in (a), while in (b) it moves from vacuum to SF and crosses no critical points. The signatures in $R$ of each state, identified in Figs.~\ref{highT} and \ref{phasesignatures}, are visible here as a linear increase in $R$ at high $T$, a plateau in the SF and a decay to zero in the MI. 
In panel (b), note that even with no MI, the window of critical fluctuations at temperatures $T>T_c$ of the strongly coupled SF varies with $\mu/U$ because of the proximity of the  $t/U\approx0.24$ quantum critical point. 
}
\end{figure*}

\noindent {\bf Onset of Quantum Effects $T^\ast$:} The deviation of $R^{-1}$ from linear $T$ behavior
defines the temperature $T^*$ at which quantum effects first become evident. 
For bosonic systems at low density $n\lesssim0.1$ we observe bunching (see supplement Fig.~1),
an enhancement of occupancy at a site, due to quantum statistics and 
manifested as $R>\beta$.

At higher density, the bunching tendency from statistics competes with inter-boson repulsion that tends to keep bosons apart.
In this regime, density correlations between sites are anti-correlated (see supplement Fig.~2).
It can be shown in general that 
$
R = {\kappa_i}/{\delta n_i^2}=\beta\left(1+\left [ \sum_{i\neq j}\left(\eone{n_in_j}-\eone{n_i}\eone{n_j}\right)\right ]/(V\delta n_i^2)\right)$
which implies that for anti-correlated density fluctuations, $R<\beta$ as seen in Fig.~\ref{highT}.

\noindent {\bf Phases:} Understanding the behavior of $\kappa_i$ and $\delta n_i^2$, the observables composing the LFD ratio, is essential for understanding the signatures of the phases in $R$ (see Fig.~\ref{phasesignatures} for typical MI and SF systems).  
The peak in $R$ at a temperature $T_{max}$ is a generic feature of the system entering an ordered phase.

{\underline {Superfluid $T_c$:}} The gapless collective excitations in the SF cause both $\kappa_i$ and $\delta n_i^2$ to approach a constant value as $T\rightarrow 0$. As the system condenses at $T_c$, the compressibility $\kappa_i$, a long wavelength response function,
exhibits a peak because of critical fluctuations. Since the appearance of long-range phase coherence does not affect the smoothly decreasing local $\delta n_i^2$, the peak in $R$ mirrors the peak in $\kappa_i$ near $T_c$. Comparison with the SF $T_c$ calculated directly from 
our QMC simulations, determined by the vanishing of the superfluid density, confirms that $T_{max}\approx T_c$, illustrated in Fig.~\ref{phasesignatures}(d). 

{\underline {Mott Insulator $T_\Delta$:}} The Mott gap suppresses the low energy excitations contributing to $\kappa_i$, causing $\kappa_i$ to vanish as $T\rightarrow 0$. On the other hand, in spite of the Mott gap, the local number fluctuations remain finite down to the lowest temperatures because of local quantum fluctuations. As the temperature is increased,
in contrast to the SF-normal phase transition, the MI crosses over into the normal state with no transition and therefore shows no specific signature in $\kappa$. Thus, the peaks of $R$ and $\kappa_i$ do not necessarily line up. We identify the peak in $R$ with the MI crossover temperature $T_\Delta$ as confirmed again by QMC simulations, where $T_{\Delta}$ corresponds to vanishingly small compressibility (see Fig.~\ref{phasediagram}(a,b)).
We further determine the zero temperature energy gap $\Delta_{ph}$ to add a particle or a hole, whichever is smaller~\cite{fisher89}
by fitting $\kappa_i$ or $R$ by $e^{-\beta\Delta_{ph}}$ within the MI. The extracted $\Delta_{ph}$ agrees very well with QMC simulations\cite{sansone08} deep in the MI (Fig.~\ref{phasesignatures}(e)), but differ on approaching the critical $(\mu/U)_c$ points, which we attribute mainly to finite temperature and size effects.

We put together our knowledge of the signatures in $R$ in the various quantum and classical phases to construct the phase diagram in Figs.~\ref{phasediagram}(a,b) at couplings above and below $(t/U)_c$. The density in each diagram changes from vacuum to $n\approx1.5$ particles per site.

\noindent {\bf Critical Regime:}
The degeneracy temperature $T^*$ depends on $t/U$ and $\mu/U$ and is reassuringly independent of the underlying critical points (see Fig.~\ref{phasediagram}(a)).
In the temperature range between $T^*$ and $T_{max}$, quantum critical fluctuations lead to
$R\propto\beta$ with a slope distinctly less than unity (Fig.~\ref{phasediagram}(e)), and $\kappa_i$ and $\delta n_i^2$ also display nontrivial behavior.
This region is clearly largest near the critical points in Fig.~\ref{phasediagram}(a), but the proximity of the critical coupling $(t/U)_c\approx0.24$ opens up a quantum critical region at intermediate density and temperature in Fig.~\ref{phasediagram}(b) as well. 
For both $t/U$ couplings, the window of critical behavior narrows as the density increases, reflecting a change from quantum criticality (proximity to quantum critical point) to classical criticality (as expected near the SF-normal phase transition).

In conclusion, while both the thermodynamic $\partial N /\partial\mu=\beta\;\delta N^2$ and the local $\partial n_i / \partial \mu_i=\beta\;\delta n_i^2$ FDT are exact relations useful for estimating the temperature,
what is new in our proposal is the construction of a LFD ratio $R$ defined in equation~(\ref{definitionR}), that involves measuring local density fluctuations in response to a global chemical potential change.
We show that $R$ is sensitive to far more than just the temperature.
As proof of principle, we have demonstrated that $R$ identifies phases and critical regimes as well as estimates the Mott mobility gap $\Delta_{ph}$. 
It should therefore be possible to {\em experimentally} map out finite temperature phase diagrams, as in Fig.~\ref{phasediagram}(a,b), without the need for individualized QMC simulations.
It is also possible to extend $R$ to other quantities such as probing spin susceptibility and corresponding spin fluctuations for magnetic systems. 
Given the very fundamental basis on which the LFD ratio is constructed, we expect it to be an ideal candidate for probing phases and quantum criticality of general quantum Hamiltonians.

{\bf Acknowledgments}

This work was financially supported by the NSF DMR-0907275 and ICAM (ED), ARO
W911NF-08-1-0338 (NT) and the DARPA OLE program.

\end{document}


\title{Supplementary Information}
\maketitle

\section{Bunching or Anti-Bunching}

\begin{figure}[h]
\begin{center}
\includegraphics[width=0.4\textwidth]{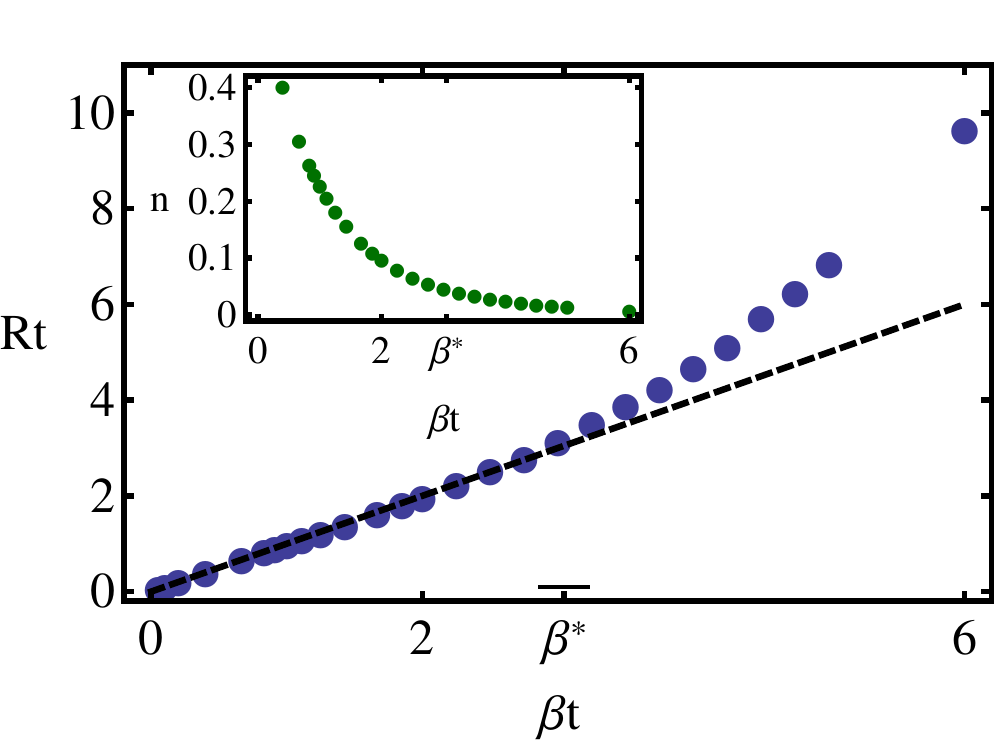}
\vspace{-0.5cm}
\end{center}
\caption{Boson Bunching due to Statistics at Low Density. \label{bunching} The main plot shows $R$ (circles) becomes larger than $\beta$ (dashed line) at low temperatures. The inset plots the density (circles) over the same temperature range. The parameters are $t/U=0.15$, $\mu/U=-0.2$ and the error in $\beta^*$ is indicated by the horizontal bar. Note that this system does not become superfluid at the temperatures probed.
}
\end{figure}

The quantity $R$ depends on the ratio of the nonlocal number fluctuations, $\delta n_i \delta n_j=\eone{n_i n_j}-\eone{n_i}\eone{n_j}$ between sites $i$ and $j$, to the local number fluctuations $\delta n_i^2=\delta n_i \delta n_i$. Specifically,
\begin{equation}
R=\frac{\kappa_i}{\delta n_i^2}=\beta\left(1+\frac{\sum_{j\neq i}\delta n_i \delta n_j}{V\delta n_i^2}\right). \label{Req}
\end{equation}
At high temperatures, the nonlocal number fluctuations are vanishingly small and $R\approx\beta$. Any finite nonlocal number fluctuations cause $R$ to deviate from $\beta$. Boson statistics favor bunching, or positively correlated nonlocal number fluctuations, and this bunching behavior is observed in $R$ in the Bose-Hubbard model at very low densities. Bunching implies the second term in equation (\ref{Req}) is positive and $R>\beta$, as shown in Fig. \ref{bunching}. 

\begin{figure}[h!]
\begin{center}
\includegraphics[width=0.4\textwidth]{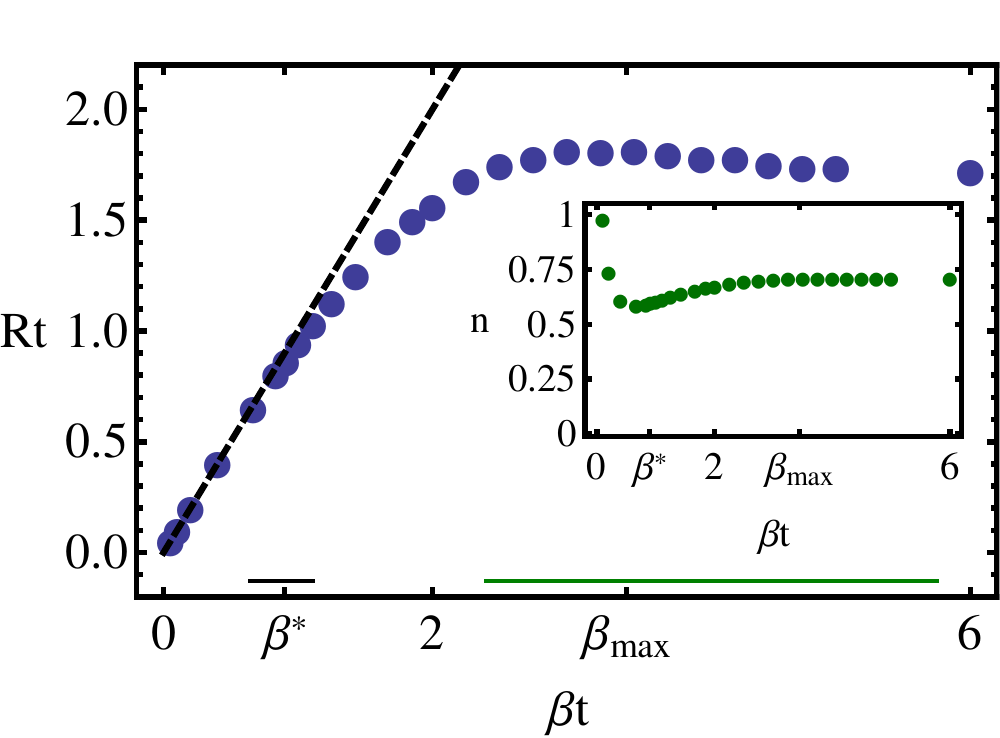}
\vspace{-0.5cm}
\end{center}
\caption{Interactions cause Bosons to Anti-Bunch at Moderate Densities. \label{antibunching} The main plot shows $R$ (circles) becomes smaller than $\beta$ (dashed line) at low temperatures when the system enters a quantum phase (superfluid in this case). The large error in $\beta_{max}$ (horizontal bar) is characteristic of the superfluid in the presence of large interactions, where the peak in $R$ is difficult to resolve. The inset shows density (circles) as a function of $\beta$. The parameters are $t/U=0.15$, $\mu/U=0.05$.
}
\end{figure}

Anti-bunching arises from the repulsive interactions in the BHM. Interactions energetically favor negatively correlated nonlocal number fluctuations and overwhelm the statistical tendency to bunch at finite density ($\gtrsim0.1$) and at (t/U) near the critical $(t/U)_c$. This implies the second term in equation (\ref{Req}) is negative and $R<\beta$ at low temperatures. This is observed to be the case in any quantum phase near $(t/U)_c$ and is exemplified in Fig. \ref{antibunching}.

\newpage

\section{The Particle-Hole Gap}

\begin{figure}[h]
\begin{center}
\includegraphics[width=0.4\textwidth]{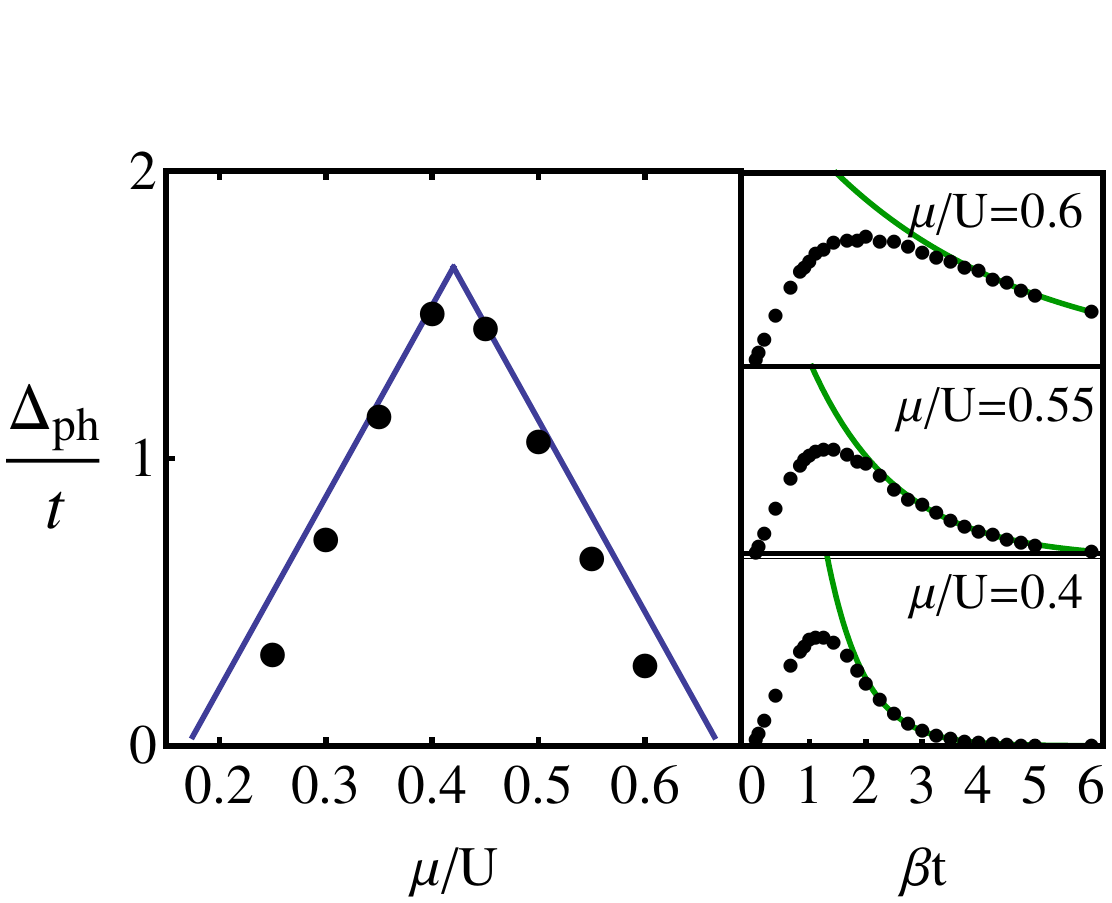}
\vspace{-0.5cm}
\end{center}
\caption{Extracting the Particle-Hole Gap from $R$. \label{phgap} The righthand column shows $R$ (circles) at various mu/U values and the exponential fit to extract $\Delta_{ph}$ (solid line). Plotted in the lefthand figure is the extracted $\Delta_{ph}$ (circles) and the results from $T=0$ QMC calculations by Capogrosso-Sansone, {\it et al.}
}
\end{figure}

The particle-hole gap $\Delta_{ph}$ is the minimal energy required to either insert or remove a particle from a Mott state. This energy scale can be extracted from $R$ by fitting its decay from the peak to zero with the simple exponential form $e^{-\beta\Delta_{ph}}$. This form also describes the decay of the compressibility $\kappa_i$ as a function of $\beta$ within the Mott state. We compare extracted $\Delta_{ph}$ with $T=0$ QMC results from Capogrosso-Sansone, {\it et al.}, in Fig. \ref{phgap}.

\section{The Fluctuation-Dissipation Theorem}
The fluctuation-dissipation theorem (FDT) relates the imaginary part of the generalized susceptibility to the correlation function. For completeness, we discuss (a) the salient points in the derivation of the quantum FDT; (b) the static structure factor; (c) the high temperature limit of the quantum FDT and its reduction to the classical FDT. (d) Further, we show that for a conserved quantity, the classical FDT expression is valid even in the quantum regime, which may seem somewhat surprising. And lastly, as a test, we provide the compressibility and the number fluctuations of the ideal gas in the high temperature limit.

\subsection{The Quantum Fluctuation-Dissipation Theorem}

Consider a quantum system of volume $\Omega$ defined by a time-independent Hamiltonian $\hat{H}$ with many body states and energies $\hat{H}\ket{\Psi_n}=\epsilon_n\ket{\Psi_n}$, 
perturbed by a probe $\hat{H}^\prime(t)$. We assume that the external perturbation $F_A(t)$ couples to an operator $\hat A$ of the system via
$\hat{H}^\prime(t)=-\hat{A}F_A(t)$. For  spatially varying probe $\hat{H}^\prime(t)=-\int d{\bf r}\hat{A}({\bf r}) F_A({\bf r}, t )=-{1\over \Omega} \sum_{\bf q} \hat{A}_{-\bf q}F_{\bf q} (t)$. For example, in our case 
$\hat{A}({\bf r})$ is the density operator
${\hat n}({\bf r})={\hat \Psi}^\dagger({\bf r})  {\hat \Psi}({\bf r})={1\over \Omega}\sum_{\bf q} e^{i {\bf q}\cdot{\bf r}}
{\hat n}({\bf q})$ where ${\hat n}({\bf q})=\sum_{\bf k}{\hat a}^\dagger_{{\bf k}-{\bf q}/2} {\hat a}_{{\bf k}+{\bf q}/2}$.

The response $\langle {\hat B}\rangle$ to linear order in the perturbation is 
$\langle {\hat B}\rangle (t) = \int_{-\infty}^{\infty} dt^\prime \chi_{BA}(t-t^\prime)F_A(t^\prime)$
where $\chi_{BA}(t-t')=i\theta(t-t')\eone{\left[\hat{B}(t),\hat{A}(t^\prime)\right]}$. 
By using a spectral representation in terms of exact eigenstates of $\hat{H}$
and the Heisenberg representation of the time dependent operators, we obtain
\begin{equation}
\chi_{BA}({\bf q},\omega)=
\frac{1}{\Omega}\sum_{m,n}\frac{e^{-\beta\epsilon_m}}{\mathcal{Z}}\left[
{     {(A_{\bf -q})_{mn}(B_{\bf q})_{nm}}  \over {\omega+i\eta + \epsilon_{nm}} }   - 
{     {(B_{\bf q})_{mn}(A_{\bf -q})_{nm}}   \over {\omega+i\eta - \epsilon_{nm}} }
\right]
\label{fullchi}
\end{equation}

where $\epsilon_{nm}=\epsilon_n-\epsilon_m$ and $\eta=0^+$ is a small positive number to ensure convergence as $t\rightarrow \infty$. 
The well-known identity $\displaystyle\lim_{\eta\rightarrow 0^+} {{1}\over {x\pm i\eta}} = P({1\over x}) \pm i\pi \delta (x)$ yields the imaginary part of the response function,
\begin{equation}
\chi''_{BA}({\bf q},\omega)=\frac{\pi}{\Omega}\sum_{m,n}\frac{e^{-\beta\omega_m}}{\mathcal{Z}}\left[(A_{-{\bf q}})_{mn}(B_{{\bf q}})_{nm}\delta(\omega-\epsilon_{nm})-(B_{{\bf q}})_{mn}(A_{-{\bf q}})_{nm}\delta(\omega+\epsilon_{nm})\right].
\label{imchi}
\end{equation}

Next we consider the corresponding correlation function typically measured in a scattering experiment defined by
\begin{flalign}
S_{BA}({\bf r},t; {\bf r}^\prime, t^\prime) & =\eone{\hat{B}({\bf r}, t)\hat{A}({\bf r}^\prime, t^\prime)}\\
S_{BA}({\bf q},t-t^\prime) & =\frac{1}{\Omega}\eone{\hat{B}_{\bf q}(t)\hat{A}_{-{\bf q}}( t^\prime)}
\label{correlation}
\end{flalign}
for a translationally invariant system. 
Using the spectral representation, we obtain
\begin{equation}
S_{BA}({\bf q},\omega)=\frac{2\pi}{\Omega}\sum_{m,n}\frac{e^{-\beta\epsilon_m}}{\mathcal{Z}}(B_{\bf q})_{mn}(A_{-{\bf q}})_{nm}\delta(\omega-\epsilon_{nm}).
\label{corr2}
\end{equation}

By exchanging the indices in the second term in Eq.\ref{imchi}, we obtain the quantum fluctuation dissipation theorem (QFDT)
\begin{flalign}
\chi''_{BA}({\bf q},\omega)&=\frac{\pi}{\Omega}(1-e^{-\beta\omega})\sum_{m,n}\frac{e^{-\beta\omega_m}}{\mathcal{Z}}(B_{\bf q})_{mn}(A_{-{\bf q}})_{nm}\delta(\omega-\omega_{nm})\\
&=\frac{1-e^{-\beta\omega}}{2}S_{BA}({\bf q},\omega).
\label{QFDT}
\end{flalign}

\subsubsection{Static Structure Factor}

The static structure factor is defined by
\begin{flalign}
S_{BA}({\bf q}) \equiv S_{BA}({\bf q},t=0)&=\int_{-\infty}^{\infty}\frac{d\omega}{2\pi} S_{BA}({\bf q},\omega)\\
&=\int_{-\infty}^{\infty}\frac{d\omega}{2\pi}\frac{2}{1-e^{-\beta\omega}}\chi''_{BA}({\bf q},\omega).
\end{flalign}
Using the oddness property $\chi''_{BA}(-\omega)=-\chi''_{BA}(\omega)$ yields
\begin{equation}
S_{BA}({\bf q})={1\over \Omega}\eone{\hat{B}_{\bf q}\hat{A}_{-{\bf q}}} =\int_{0}^{\infty}\frac{d\omega}{\pi}\coth\left(\frac{\beta\omega}{2}\right)\chi''_{BA}({\bf q},\omega).
\end{equation}

\subsubsection{The High Temperature Limit of the QFDT}

At temperatures $k_BT\gg \hbar \omega$ larger than any characteristic frequencies of the system, 
$\coth\left(\frac{\beta\omega}{2}\right)\rightarrow 2/\beta\omega$
and the static structure factor reduces to
\begin{flalign}
S_{BA}({\bf q})&=\frac{2}{\beta}\int_{0}^{\infty}\frac{d\omega}{\pi}\frac{\chi''_{BA}({\bf q},\omega)}{\omega}\\
&=k_B T \chi'_{BA}({\bf q},\omega=0)
\label{highTlimit}
\end{flalign}
where we have used the Kramers-Kr\"onig relation $\chi^\prime_{BA}({\bf q},\omega)=P\int_{-\infty}^{\infty}\frac{d\omega^\prime}{\pi}\frac{\chi''_{BA}(q,\omega^\prime)}{\omega^\prime -\omega}$ to relate the real and imaginary parts of the response function. 
Since at $\omega=0$ the imaginary part is zero, we can replace $\chi^\prime$ by simply $\chi$.

\subsection{The QFDT for Conserved Quantities}

We next use (i) the definition of the correlation function for a conserved quantity, (ii) the quantum FDT, and (iii) Kramers-Kr\"onig relation to finally derive
$\chi_{AA}(q\rightarrow0,\omega=0)=\beta S_{AA}(q\rightarrow0,t=0)$. The derivation is detailed below.

A conserved quantity ${\hat A}({\bf q=0})\equiv A_0$ such as 
${\hat n}({\bf q=0})=\sum_{\bf k}{\hat a}^\dagger_{\bf k} {\hat a}_{\bf k}=N$, the total number of particles, 
commutes with the Hamiltonian, $\left[\hat{A}_0,\hat{H}\right]=0$. 
This implies that the matrix element $\langle \Psi_m \mid \left[\hat{A}_0,\hat{H}\right] \mid \Psi_n\rangle =0$ or equivalently $(\epsilon_n-\epsilon_m) (A_0)_{m,n}=0$. If $m\neq n$, we must have 
$(A_0)_{m,n}=0$ which results in 
\begin{equation}
\lim_{\omega\rightarrow 0} \lim_{{\bf q}\rightarrow 0} 
\chi_{AA}({\bf q},\omega)=
\lim_{\omega\rightarrow 0}
\frac{2}{\Omega}\sum_{m,n}\frac{e^{-\beta\epsilon_m}}{\mathcal{Z}}\left[
     \frac{\epsilon_{nm} \mid (A_0)_{mn}\mid^2}{(\omega+i\eta)^2 - \epsilon^2_{nm}}\right]   =0
\end{equation}
using Eq. \ref{fullchi}.
Thus $ \chi_{AA}({\bf q}=0,\omega\rightarrow 0)=0$ if $A({\bf q}=0)$ is a conserved quantity.

However, the situation is totally different if we change the order of limits.
\begin{equation}
\lim_{{\bf q}\rightarrow 0} \lim_{\omega\rightarrow 0} 
\chi_{AA}({\bf q},\omega)=
\lim_{{\bf q}\rightarrow 0}
\frac{2}{\Omega}\sum_{m,n}\frac{e^{-\beta\epsilon_m}}{\mathcal{Z}}\left[
{     {  \epsilon_{nm} \mid (A_{\bf q})_{mn}\mid^2}  \over {(\omega+i\eta)^2 - \epsilon^2_{nm} } } \right] = n^2 \kappa
\label{above}
\end{equation}
For density operators, the last equality in Eq.~\ref{above} follows from the perturbation
${\hat H}^\prime=-\int d{\bf r} \delta {\hat n}({\bf r}) \delta {\hat \mu}({\bf r},t) =-{1\over \Omega} \sum_{\bf q} \delta {\hat n}_{-{\bf q}} \delta {\hat \mu}_{\bf q}(t)$
which produces a response in the system 
 \begin{flalign}
 \chi_{nn}({\bf q}\rightarrow 0,\omega=0) & ={ {\eone{\delta {\hat n}_{-{\bf q}\rightarrow 0} }} \over { \delta {\hat \mu}}_{{\bf q}\rightarrow 0} }  (\omega=0) \\
 &  = \left ({{\partial n} \over {\partial \mu}}\right )_{T,\Omega} = n^2 \kappa
 \end{flalign}
 where $\kappa$ is the isothermal compressibility. 

Another way to understand the behavior of QFDT for a conserved quantity $\eone{\hat{A}({\bf q=0},t)}$ is to note that it is independent of $t$.
In addition the correlator 
$\eone{\hat{A}({\bf q=0},t)\hat{A} ({-{\bf q}=0},t^\prime)}$ is independent of $t-t^\prime$ and hence its Fourier transform must be a delta function in frequency.

Thus from Eq.~\ref{correlation}
\begin{equation}
S_{AA}({\bf q}=0,\omega)= 2 \pi \delta(\omega) S_{AA}({\bf q}=0)
\label{S-cons}
\end{equation}

From the quantum FDT Eq.~\ref{QFDT} we obtain for a conserved quantity whose correlation function is given by Eq.~\ref{S-cons},
\begin{equation}
\chi''_{AA}({\bf q}=0,\omega)=(1-e^{-\beta\omega}) \pi S_{AA}({\bf q}=0) \delta(\omega)
\end{equation}
By using the Kramers-Kr\"onig relation we get
\begin{flalign}
\chi_{AA}({\bf q}=0,\omega\rightarrow 0) & = S_{AA}({\bf q}=0) \int_{-\infty}^{\infty} d \omega {{1-e^{-\beta\omega}}\over {\omega}} 
\delta(\omega)\\
& = \beta S_{AA}({\bf q}=0) ={\beta \over \Omega}\eone{\hat{A}^2({\bf q}=0)} 
\end{flalign}
For density operators we thus get 

\begin{equation}
\chi_{nn}({\bf q}=0) = {\beta \over \Omega}\eone{\hat{n}^2({\bf q}=0)} = n^2\kappa
\end{equation}
We would like to stress that while this result may look similar to the high temperature limit of equation (\ref{highTlimit}), it is valid in the quantum regime for a conserved quantity.

\subsection{The Ideal Gas}

We derive the high temperature behavior of the compressibility, $\partial n/\partial\mu$, and number fluctuations $\delta N^2\equiv\langle N^2\rangle - \langle N\rangle^2$
 of the ideal gas. From the equation of state $PV=Nk_BT$,
we get 
\begin{equation}
\kappa_T\equiv-\frac{1}{V}\thone{V}{P}{T,N} \equiv {1\over n^2} \frac{\partial n}{\partial\mu}= \frac{\beta}{n}
\end{equation}
where $n=N/V$.

For an ideal gas, the chemical potential is related to the density by $\beta \mu= -\;{\rm log}\left(\frac{1}{n\lambda_T^d}\right)$, where the thermal deBroglie 
wavelength $\lambda_T=h/\sqrt{2\pi mk_BT}$. For a fixed $\mu$, the high temperature expansion of $n(\mu,T)$ is
\begin{equation}
n=\frac{e^{\beta\mu}}{\lambda_T^d}\sim T^{d/2}\left(1+\frac{\mu}{k_BT}+\frac{1}{2}\left(\frac{\mu}{k_BT}\right)^2\right).
\end{equation}
which implies that the temperature dependence of the local compressibility is 
\begin{equation}
n^2\kappa=\frac{\partial n}{\partial\mu}\sim T^{d/2-1}\left(1+\mu\;T^{-1}\right).
\end{equation}

Using the local fluctuation-dissipation theorem at high temperatures, we find
\begin{equation}
\eone{\delta n^2}\approx\frac{\partial n}{\partial\mu} T\sim T^{d/2}\left(1+\mu T^{-1}\right),
\end{equation}
as we might have guessed from a simple Brownian motion or diffusion model of the number fluctuations. 

Thus, we find that in 2D, while the local compressibility $n^2\kappa$ is independent of temperature (note that in the Letter we have absorbed the $n^2$ factor in the definition of $\kappa$)
and the number fluctuations $\eone{\delta n^2}$ scale linearly with $T$ . 

In the presence of interactions, both number fluctuations and compressibility deviate from their classical values, however their ratio $R=n^2\kappa/\eone{\delta n^2}$ continues to be linear in $T^{-1}$
as determined by the fluctuation-dissiaption theorem (see Fig.~2 of Letter).  Only when quantum effects become important does $R$ exhibit non-trivial behavior.